\documentclass{appolb}
\usepackage{graphicx}

\begin{document}
\title{Does the Gogny interaction need a third Gaussian?%
\thanks{Presented at Zakopane Conference on  Nuclear Physics 2016}%
}
\author{D. Davesne,P. Becker
\address{Universit\'e de Lyon, Universit\'e Lyon 1, CNRS/IN2P3, Institut de Physique Nucl{\'e}aire de Lyon, UMR 5822, F-69622 Villeurbanne cedex, France}
\\
{A. Pastore
}
\address{Department of Physics, University of York, Heslington, York, Y010 5DD, United Kingdom}\\
{J. Navarro
}
\address{IFIC (CSIC-Universidad de Valencia), Apartado Postal 22085, E-46.071-Valencia, Spain}
}
\maketitle
\begin{abstract}
By considering infinite matter constraints only, we suggest in this paper that the Gogny interaction should benefit from a third Gaussian in its central part. A statistical analysis is given to select the possible ranges which are compatible with these constraints and which minimize a $\chi^2$ function.
\end{abstract}
  
\section{Introduction}

The nuclear energy density functional theory (NEDF) is the tool of choice for the description of nuclear properties from drip-line to drip-line and from light to super-heavy elements~\cite{ben03}. 
The quest of the universal functional represents one of the major current challenges in low-energy nuclear physics as testified by the large scientific collaborations as UNEDF~\cite{ber07}. 

To build a functional, a possible framework is to consider only the non-relativistic limit and the functionals which can be derived from an effective interaction~\cite{rai11}. By limiting ourselves to these two simplifications, we can identify three main families of functionals: the ones derived either from the zero-range (Skyrme~\cite{sky59,vau72}) or the finite-range (Gogny~\cite{gog75,dec80} or M3Y~\cite{ber77,nak03}) interactions. 

In a recent article~\cite{dav16ANNALS}, we have shown that the zero-range Skyrme interaction can be considered as a low-momentum expansion of the finite range one. By considering the partial-wave expansion method~\cite{dav15s,Dav16PRC}, we have shown that while formally a finite-range interaction gets contributions from \emph{all} partial waves, in reality by restricting to $S,P,D$ and ultimately $F$ partial waves, we can obtain an excellent description of some basic infinite matter properties (INM) as the equation of state (EoS).
This analysis motivated us to perform new studies on the \emph{extended} Skyrme interactions~\cite{car08,rai11}, showing that the N3LO version, $i.e.$ including explicitly terms up to $6$th order in gradients, is able to grasp the main physical properties of a more complex Brueckner-Hartree-Fock (BHF) calculation~\cite{bal97}. In Ref.~\cite{Dav15AA}, we have derived the parameters of the N3LO EoS to be used in astrophysical simulations. One of the major achievements of introducing higher order gradient terms in the Skyrme functional was the additional flexibility required to reproduce the spin (S)-isospin (T) decomposition of the EoS~\cite{les07}.

The ST-decomposition of the EoS is routinely used as a constraint during the fitting procedure of several effective interactions~\cite{nak03,cha07,cha15}: for example the (S=0,T=1) channel is a useful constraint to determine the pairing properties of the efective interaction. Although these (S,T) channels are not strictly speaking observables, the results obtained by different many-body calculations all agree on the sign of the EoS in the different channels~\cite{Dav15} and on the trend as a function of the density of the system.

Despite their importance, we have noticed in Ref.~\cite{dav16ANNALS} that it is very difficult to get a satisfactory reproduction of these channels by using a standard Gogny interaction~\cite{dec80}: even a re-fit of the parameters that doesn't take into account finite nuclei constraints doesn't improve significantly the symmetric nuclear matter (SNM) properties. In the present article, we therefore discuss the possibility of modifying the Gogny interaction by adding a third range.

The article is organised as follows: in Sec.~\ref{sec:inm} we recall the main INM properties of the Gogny interaction, while in Sec.\ref{sec:3g} we discuss the role of a third range on the Gogny interaction. Our conclusions are presented in Sec.\ref{sec:conc}.

\section{Symmetric nuclear matter equation of state}\label{sec:inm}

The effective Gogny interaction was proposed in the seventies aiming at offering a fair description of static properties of spherical as well as deformed nuclei. In its standard form~\cite{dec80}, it is a sum of central $v^C$, spin-orbit $v^{LS}$ and density-dependent $v^{DD}$ terms. The inclusion of an explicit tensor term $v^T$ has been considered more recently in Refs.~\cite{Ang11,Ang12,Gra13}.
The central term consists of a sum of two Gaussian functions:
\begin{equation}\label{gog:cen}
v^C_{G}(\mathbf{r}_1,\mathbf{r}_2) = \sum_{i=1}^2 \left[ W_i+B_iP_{\sigma}-H_iP_{\tau}-M_iP_{\sigma}P_{\tau}  \right] e^{-(r / \mu^{C}_i)^2} ,
\end{equation}
where the explicit ranges values $\mu_i^C$ depend on the adopted parametrisation. Concerning the density-dependent term, it is similar to the Skyrme's one. Finally, since in INM, the tensor and the spin-orbit terms do not contribute at Hartree-Fock (HF) level to the (S,T) decomposition of the EoS~\cite{Dav16PRC}, we will not consider them explicitly in the current article.
We refer to Refs.~\cite{dav16ANNALS,ber16} for more details on their properties.

\begin{figure}[htb]
\begin{center}
\includegraphics[width=8.5cm,angle=-90]{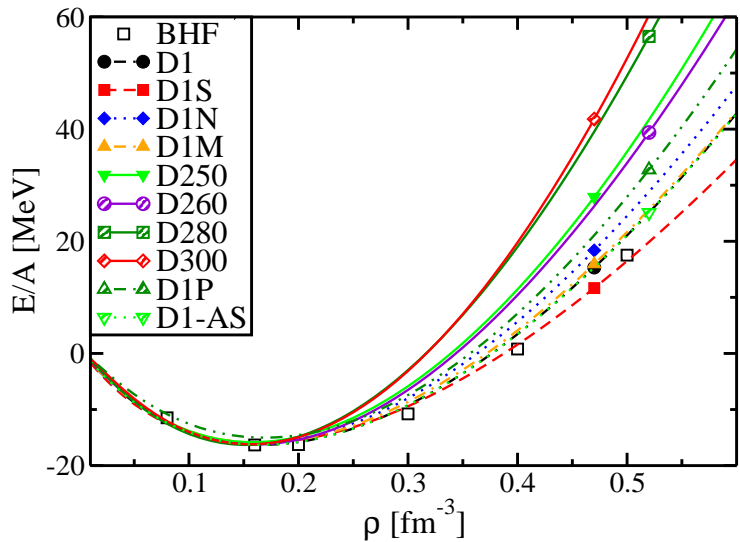}
\end{center}
\caption{(Color online) Energy per particle in SNM for different Gogny interactions as a function of the density.}
\label{eossnm}
\end{figure}

By averaging Eq.\ref{gog:cen} over HF states, we can obtain the EoS for INM. We refer the reader to Ref.~\cite{sel14} for a detailed discussion on INM properties of the Gogny interactions and also for explicit expressions. As an illustration, we show in Fig.\ref{eossnm} the SNM EoS obtained for some existing Gogny parametrisations as described in Ref.~\cite{sel14}.

\begin{figure}[htb]
\begin{center}
\includegraphics[width=8.5cm,angle=-90]{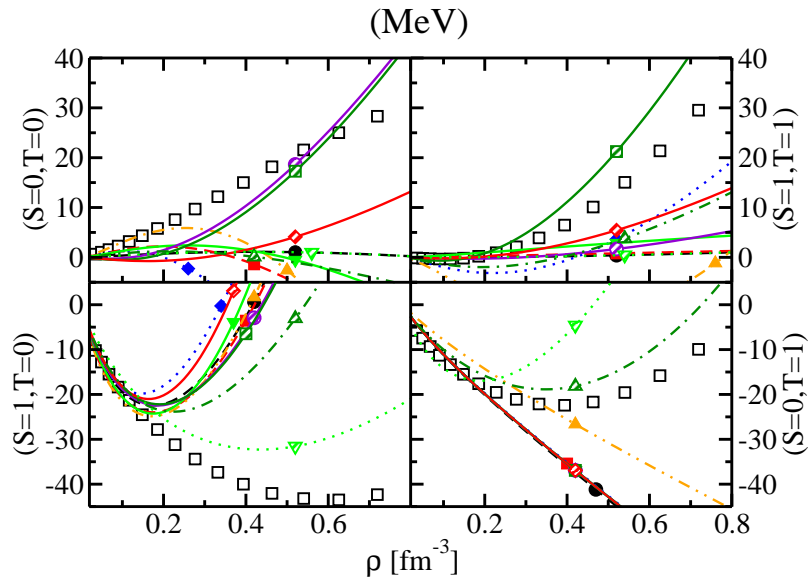}
\end{center}
\caption{(Color online) Spin-isospin decomposition of the potential energy, expressed in MeV, for various Gogny interactions. The legend is the same as in Fig.\ref{eossnm}.}
\label{ST}
\end{figure}

\noindent On the same figure we also show the BHF results. As it should be, we observe that all Gogny parametrisations reproduce fairly well the SNM properties around saturation density. On the contrary, at higher densities, some quantitatively important differences are observed. However, as discussed previously, our aim in this paper is not to comment further the EoS itself but its decomposition in the (S,T) channels. The explicit expressions for Gogny interaction have been already given in Ref.~\cite{dav16ANNALS} (for the interactions D1P and D1-AS, which have a slightly different density dependence, the results have been straigthforwardly generalised). The results are presented in Fig.\ref{ST}. None of the existing Gogny parametrisations is able to give a satisfactory representation of the ST decomposition or to get the right sign of the interaction in all channels above saturation density.

\noindent Such a result does not change if we re-fit the parameters of the interaction by taking into account only SNM properties (see Ref.~\cite{dav16ANNALS}) for details.
The additional density dependent term of D1P and D1-AS is not of much help since it explicitly enters only in the (S=1,T=0) and (S=0,T=1) channel.
We have thus decided to explore a modification of the Gogny interaction by considering an additional Gaussian.

\section{Three-Gaussian interaction}\label{sec:3g}

In this section we consider the possibility of adding a third range to the central term of the Gogny interaction, Eq.~\ref{gog:cen}.
In this exploratory work we will neglect all finite-nuclei constraints but focus only on the behavior of the EoS in SNM.
For the sake of simplicity we also freeze the long range ($\mu_1$=1.2 fm) and the power of the density dependence ($\alpha=1/3$), while we let the other two ranges as well as all the other parameters free to move.

\begin{figure}[htb]
\begin{center}
\includegraphics[width=8.5cm,angle=0]{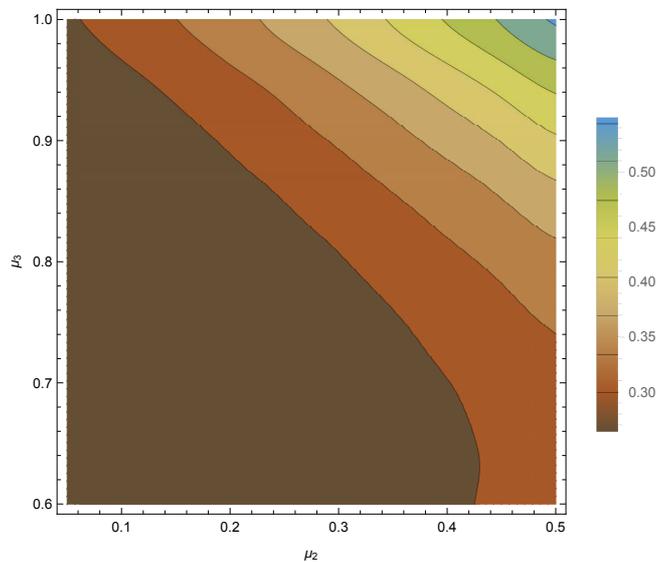}
\end{center}
\caption{(Color online) Contour lines of the $\chi^2$ function. See text for details}
\label{chi2}
\end{figure}

\begin{figure}[htb]
\begin{center}
\includegraphics[width=4.8cm,angle=-90]{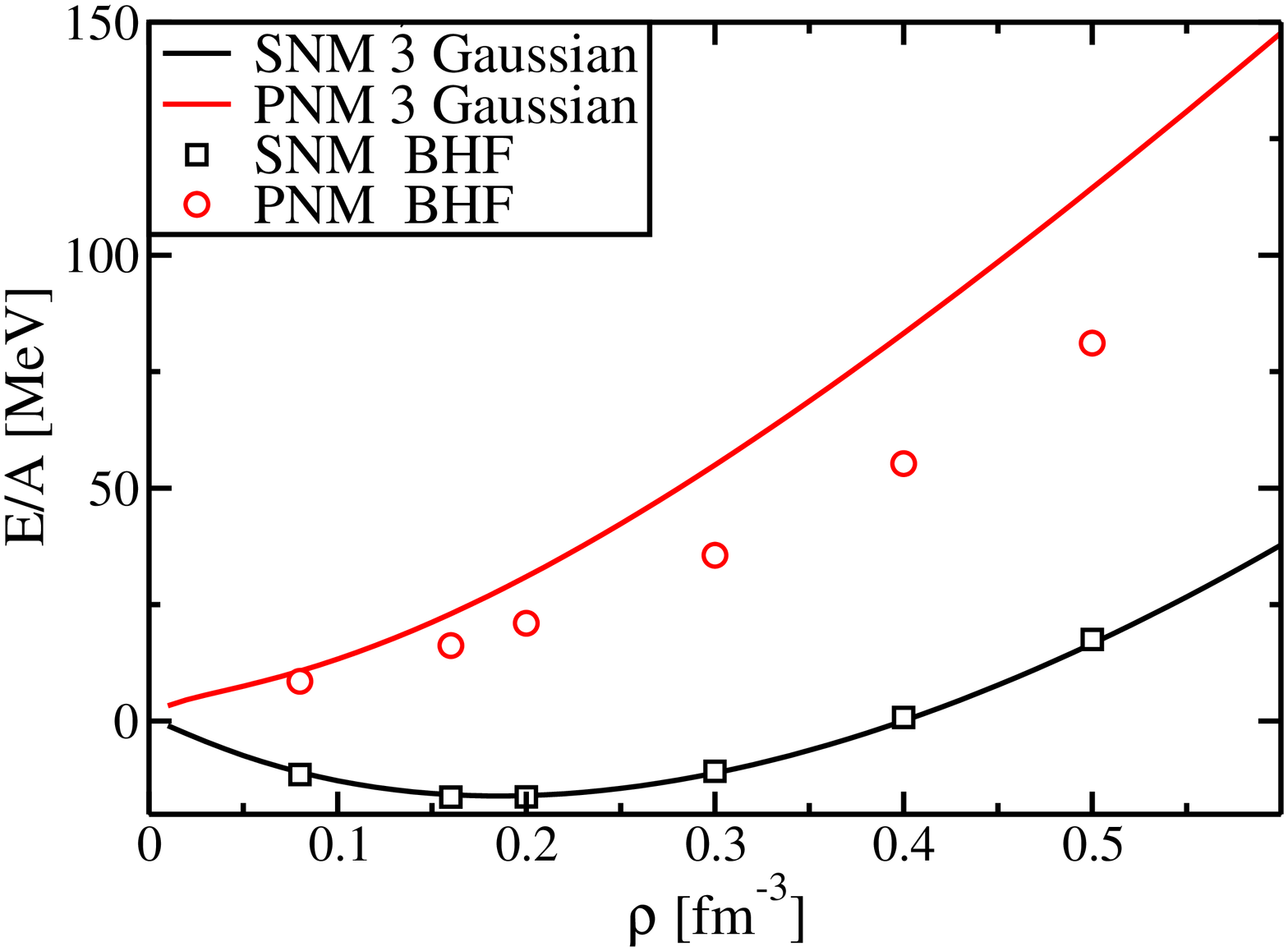}
\includegraphics[width=4.8cm,angle=-90]{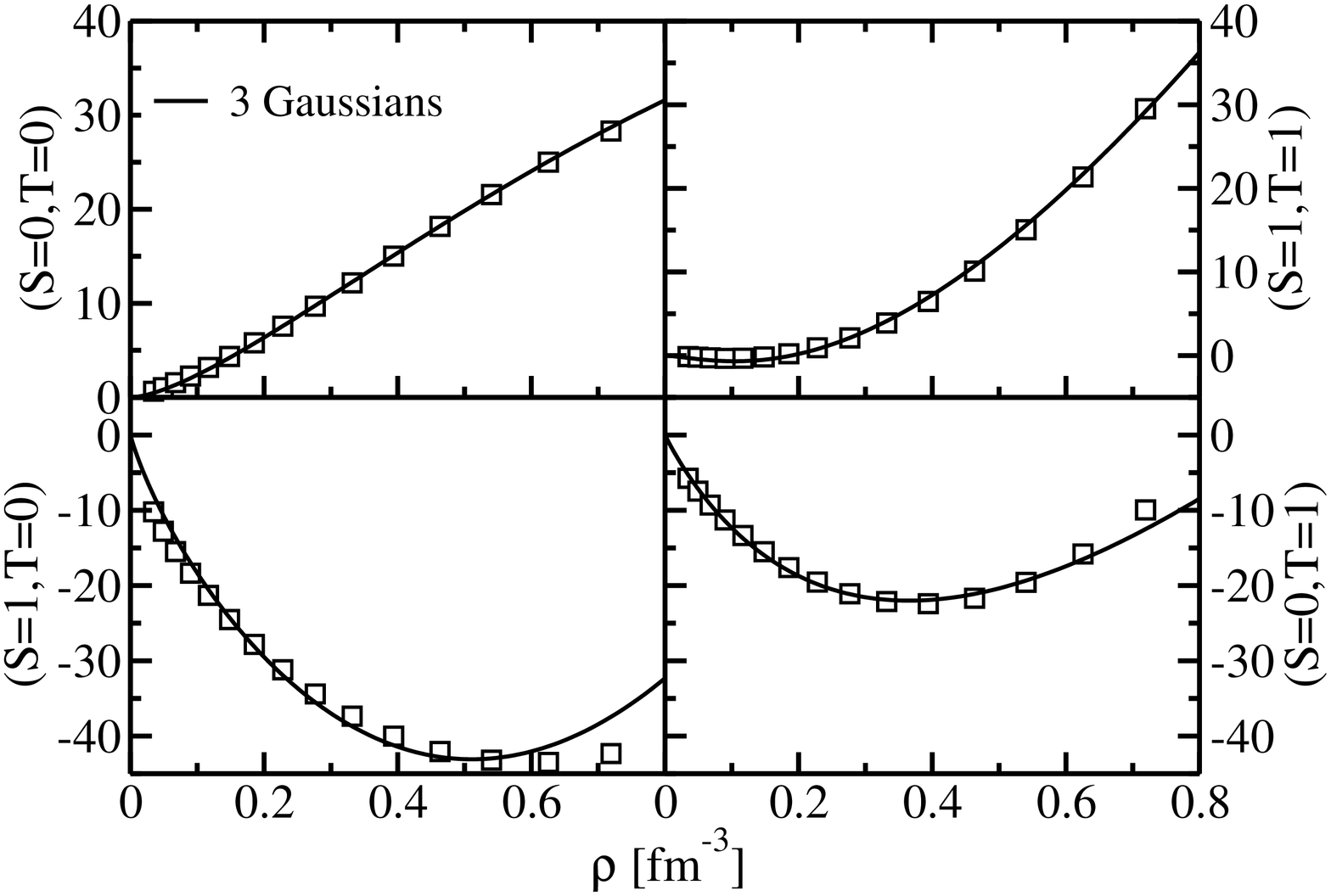}
\end{center}
\caption{(Color online) On the left panel we compare the EoS in SNM and PNM for the 3-Gaussian interaction and the BHF results~\cite{bal97}. On the right panel we compare the ST decomposition of the potential energy, expressed in MeV, for these two calculations. See text for details.}
\label{STfit}
\end{figure}

To have a first insight of the possible values for the two remaining ranges, we built a $\chi^2$ function using as \emph{observables} the BHF data on the ST channel decomposition.
In Fig.\ref{chi2}, we show the contour line of the $\chi^2$  for different values of the ranges $\mu_2,\mu_3$. We observe that the surface is quite flat thus showing that we have the freedom to chose the two ranges in a reasonable large parameter space.
For example we can choose the set ($\mu_2=0.8$ fm and $\mu_3=0.25$ fm). In Fig.\ref{STfit}, we compare the EoS for the 3-Gaussian resulting interaction with the original BHF results for SNM and also for the case of pure neutron matter (PNM). Notice that the PNM EoS does not enter in the fitting protocol.
On the right panel of Fig.\ref{STfit}, we show the detailed comparison of the ST channels.

\begin{figure}[htb]
\begin{center}
\includegraphics[width=4.8cm,angle=-90]{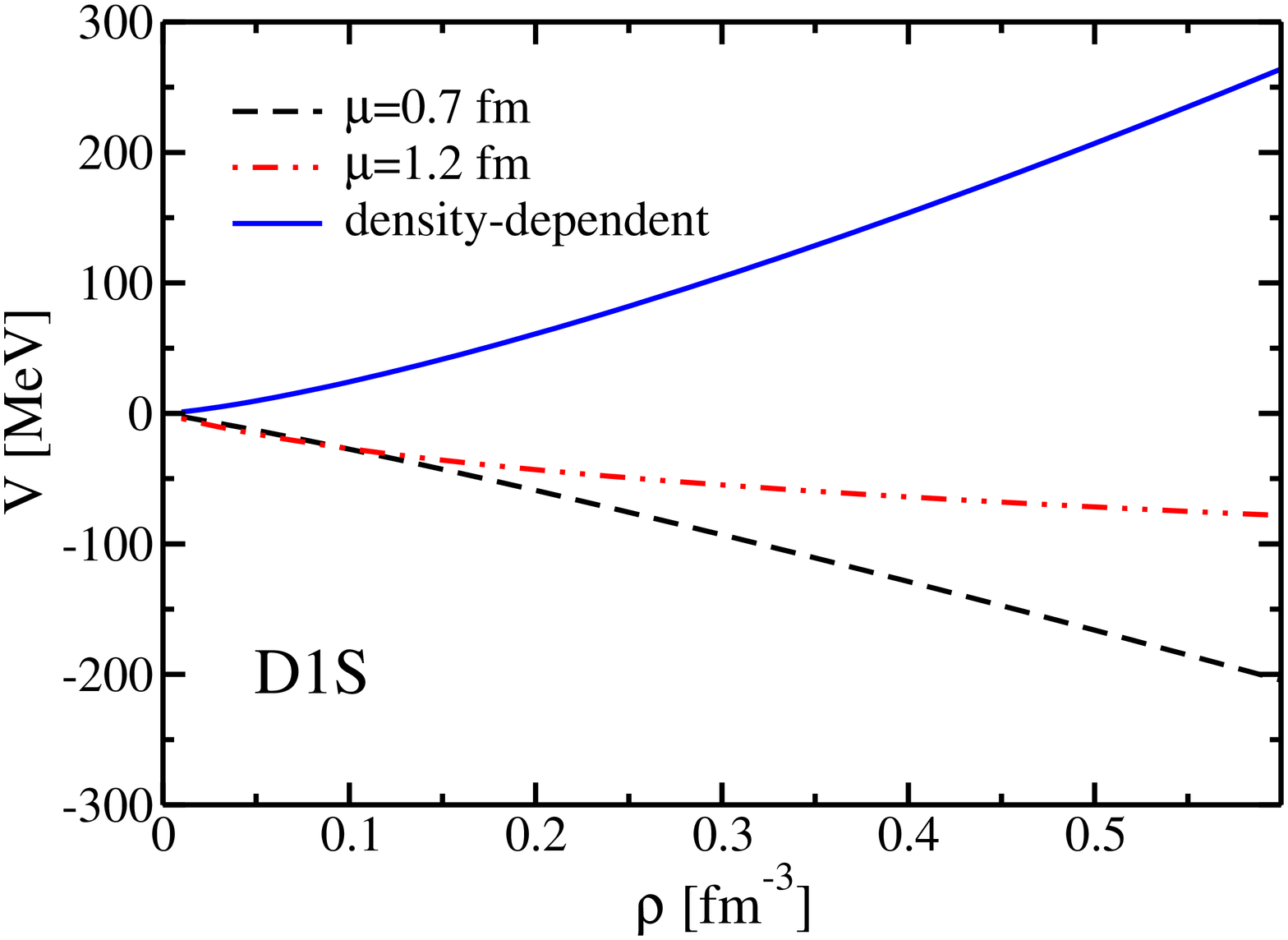}
\includegraphics[width=4.8cm,angle=-90]{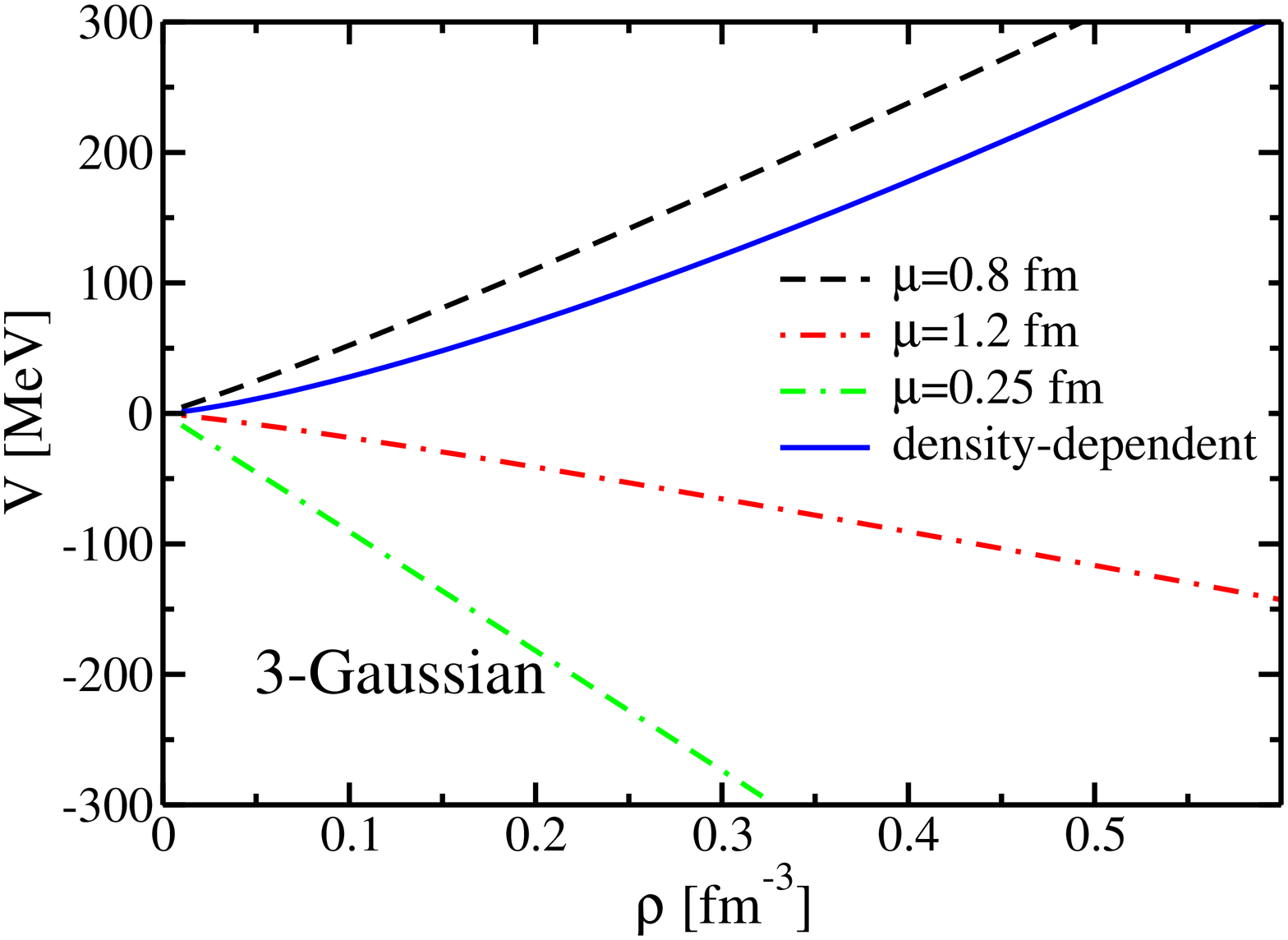}
\end{center}
\caption{(Color online) Partial contributions to the EoS in SNM for Gogny D1S (left panel) and the 3-Gaussian interaction (right panel). See text for details.}
\label{eospartial}
\end{figure}

\noindent We observe that the addition of the 3-Gaussian leads to a perfect reproduction of the channels and therefore of the EoS in SNM,
since the EoS is simply related to the (S,T) channels. In Fig.\ref{eospartial}, we show the different contributions to the total potential energy, by isolating the contribution of each range and the density dependent one. From this figure we observe that the long range part and the density dependent have roughly the same behavior in both cases. We also notice that the short range part changes sign and magnitude from D1S to 3-Gaussian and this is compensated by the strong contribution given by the third Gaussian.

\noindent We can thus conclude from this picture that the third Gaussian introduced here is not a minor correction, but gives a strong contribution to the EoS. However, focusing on BHF results for the (S,T) channels in SNM is not the end of the story, as other quantities could be modified by the addition of a third Gaussian. For instance, the value of effective mass in SNM at saturation increases from 0.73 (D1S) to 0.87, which will result in a higher density of states around the Fermi surface and larger pairing gaps. Including explicitly (S,T) results into the fitting protocol would lead to better results. Of course, more detailed investigations incorporating finite nuclei constraints are required before making definitive statements.

\section{Conclusions}\label{sec:conc}

We have investigated some properties of the Gogny interaction in the infinite medium and we have explored the possibility of adding a third Gaussian with an extra short range to improve the reproduction of Brueckner-Hartree-Fock results.
Although for the moment these results can not be considered conclusive since we have not considered explicitly properties of atomic nuclei, we think this modification could be explored in the near future since it requires no major modification in existing Gogny codes.
It is worth reminding that the improvement of the central part is not the only modification required for the Gogny interaction since for example an additional tensor term~\cite{Ang11} would also improve some aspects of the interaction.

\section*{Acknowledgments} 

We thank M. Baldo for kindly providing us with his BHF results and S. P\'eru for interesting discussions.  The work of J.N. has been supported by grant  FIS2014-51948-C2-1-P, Mineco (Spain). 

\bibliographystyle{polonica}

\bibliography{biblio}

\end{document}